\documentclass[twocolumn,showpacs,preprintnumbers,amsmath,amssymb,floatfix]{revtex4}

\usepackage{graphicx}% Include figure files
\usepackage{dcolumn}% Align table columns on decimal point
\usepackage{bm}% bold math

\begin{document}

%\preprint{APS/xxxxx}

\title{A confirmative spin-parity assignment for the key 6.15 MeV state in $^{18}$Ne of astrophysical importance}

\author{J. Hu$^1$}
\author{J.J. He$^1$}
\email{jianjunhe@impcas.ac.cn}
\author{S.W. Xu$^1$}
\author{H. Yamaguchi$^2$}
\author{K. David$^2$}
\author{P. Ma$^1$}
\author{J. Su$^3$}
\author{H. W. Wang$^4$}
\author{T. Nakao$^2$}
\author{Y. Wakabayashi$^5$}
\author{T. Teranishi$^7$}
\author{J.Y. Moon$^6$}
\author{H.S. Jung$^6$}
\author{T. Hashimoto$^8$}
\author{A. Chen$^9$}
\author{D. Irvine$^9$}
\author{S. Kubono$^{1,5}$}

\affiliation{$^1$Institute of Modern Physics (IMP), Chinese Academy of Sciences (CAS), Lanzhou 730000, China}
\affiliation{$^2$Center for Nuclear Study (CNS), University of Tokyo, Wako Branch at RIKEN, 2-1 Hirosawa, Wako, Saitama 351-0198, Japan}
\affiliation{$^3$China Institute of Atomic Energy (CIAE), P.O. Box 275(46), Beijing 102413, China}
\affiliation{$^4$Shanghai Institute of Applied Physics (SINAP), Chinese Academy of Sciences (CAS), Shanghai 201800, China}
\affiliation{$^5$RIKEN Nishina Center, 2-1 Hirosawa, Wako, Saitama 351-0198, Japan}
\affiliation{$^6$Department of Physics, Chung-Ang University, Seoul 156-756, Republic of Korea}
\affiliation{$^7$Department of Physics, Kyushu University, 6-10-1 Hakozaki, Fukuoka 812-8581, Japan}
\affiliation{$^8$RCNP, Osaka University, 10-1 Mihogaoka, Ibaraki, Osaka, 567-0047, Japan}
\affiliation{$^9$Department of Physics and Astronomy, McMaster University, Hamilton, Ontario L8S 4M1, Canada}

\date{\today}  % It is always \today, today, but any date may be explicitly specified

\begin{abstract}
Proton resonant states in $^{18}$Ne have been investigated by the resonant elastic scattering of $^{17}$F+$p$. The $^{17}$F beam was 
separated by the CNS radioactive ion beam separator (CRIB), and bombarded a thick H$_2$ gas target at 3.6 MeV/nucleon.
The recoiled light particles were measured by using three sets of ${\Delta}$E-E Si telescope at scattering angles of 
$\theta$$_{lab}$$\approx$3$^\circ$, 10$^\circ$ and 18$^\circ$, respectively. Four resonances, {\it i.e.}, at $E_{x}$=6.15, 6.30, 6.85, 
and 7.05 MeV, were observed clearly. By $R$-matrix analysis of the excitation functions, $J^{\pi}$=1$^-$ was firmly assigned to 
the 6.15 MeV state which is a key state in calculating the reaction rate of $^{14}$O($\alpha$,$p$)$^{17}$F reaction. This reaction was 
thought to be one of the most probable key reactions for the breakout from the hot-CNO cycle to the $rp$-process in type I x-ray bursts
In addition, a new excited state observed at $E_{x}$=6.85 MeV was tentatively assigned as 0$^{-}$, which could be the analog state of 
6.880 MeV, 0$^{-}$ in mirror $^{18}$O.

%The stellar reaction rate of the $^{14}$O($\alpha$, $p$)$^{17}$F reaction has been re-evaluated based on previous works.
%In addition, resonant inelastic scattering has populated three further states at excitation energies of 3.14, 3.26 and 3.95 MeV, with proton decay to
%the first excited state in $^{22}$Mg being observed. The new state at 3.95~MeV has been assigned a spin-parity of $J^{\pi}$ = (7/2$^{+}$).
%The core-excited structure of $^{23}$Al is discussed within a shell-model picture.
%the revised total reaction rate is increased by about 40\% beyond temperature $T_9=0.3$ comparing to the previous result.
\end{abstract}

% PACS codes here, in the form: \PACS code \sep code
%\keywords{Reaction induced by unstable nuclei, Decay by proton emission, nuclear astro, A}

\maketitle

% main text
Explosive hydrogen and helium burning are thought to be the main source of energy generation and a source for the nucleosynthesis of 
heavier elements in cataclysmic binary systems, such as in novae, x-ray bursts, {\it etc.}~\cite{bib:cha92,bib:wie99}. Also they 
provide an important route for nucleosynthesis of elements up to masses 100 region via the rapid-proton capture process 
(rp-process)~\cite{bib:sch01}. According to the previous estimation~\cite{bib:wie99} the $^{14}$O($\alpha$,$p$)$^{17}$F 
reaction is probably one of the key breakout reactions from the hot-CNO (HCNO) cycle to the rp-process in Type I x-ray bursts (XRBs). 
XRBs are driven by a thermonuclear runaway on the surface of an accreting neutron star, and the accretion disk has a very high density 
of $\rho$=10$^{3}$--10$^{6}$ g/cm$^{3}$. The breakout from the HCNO cycle and the onset temperature of the rp-process mainly depend 
on the cross section (or reaction rate) of the $^{14}$O($\alpha$,$p$)$^{17}$F reaction. Actually, the resonant contribution dominates 
the $^{14}$O($\alpha$,$p$)$^{17}$F reaction rate, and therefore knowledge of the resonant properties for those excited states above 
the $\alpha$ threshold ($Q_\alpha$=5.115 MeV~\cite{bib:wan12}) in compound nucleus $^{18}$Ne is required.
So far, although our understanding in the reaction rates of $^{14}$O($\alpha$,$p$)$^{17}$F has been greatly improved via, for example, 
indirect studies~\cite{bib:hah96,bib:gom01,bib:bla03,bib:cern,bib:bar10}, 
direct studies~\cite{bib:not04,bib:kub06}, as well as time-reversal studies~\cite{bib:bla01,bib:har99,bib:har02}, most of the resonant 
properties (such as, $J^{\pi}$, $\Gamma_\alpha$ and $\Gamma_p$) have not been sufficiently well determined over stellar temperatures 
achieved in XRBs. 

In temperature region below 1.0 GK, a 1$^-$ state at $E_x$=6.15 MeV dominates the $^{14}$O($\alpha$,$p$)$^{17}$F rate, while other
high-lying states dominate at high temperatures (1--3 GK) relevant for XRBs~\cite{bib:hah96}.
Thirty years ago, Wiescher {\it et al.}~\cite{bib:wie87} predicted a $J^\pi$=1$^-$ state at $E_x$=6.13 MeV in $^{18}$Ne with a width 
of $\Gamma$=$\Gamma_p$=51 keV based on a Thomas-Ehrman shift calculation. A spectroscopic factor of $C^2S$=0.03, obtained from 
the 6.198 MeV $^{18}$O analog state was adopted in their calculation. Later on, Hahn {\it et al.}~\cite{bib:hah96} found a state at 
$E_x$=6.15$\pm$0.02 MeV in both reactions of $^{16}$O($^3$He,$n$)$^{18}$Ne and $^{12}$C($^{12}$C,$^6$He)$^{18}$Ne, while not in the 
$^{20}$O($p$,$t$)$^{18}$Ne reaction. The transferred angular momentum was restricted to be $L$$\leq$2 from the measured 
($^3$He,$n$) angular distribution. Based on a Coulomb-shift calculation and Wiescher {\it et al.} prediction, a $J^\pi$=1$^-$ was 
assigned to this 6.15 MeV state. G\"{o}mez del Campo {\it et al.}~\cite{bib:gom01} studied the resonances 
in $^{18}$Ne by using the elastic scattering of $^{17}$F+$p$, and fitted the 6.15 MeV state with 1$^-$ by $R$-matrix analysis of the 
excitation function. However, the above 1$^{-}$ assignment for the 6.15 MeV state was questioned by a careful $R$-matrix 
reanalysis~\cite{bib:he10} of the previous data~\cite{bib:gom01}. If it were 1$^-$ state, He {\it et al.}~\cite{bib:he10} thought that 
the 6.15 MeV resonance should behave as a groove-like structure in the excitation function, due to the interference between the 
resonant and potential scattering. Unfortunately, our recent low-statistics measurement could not resolved this state~\cite{bib:he11}. 
Most recently, Bardayan {\it et al.}~\cite{bib:bar12} reanalyzed the elastic-scattering data in Ref.~\cite{bib:bla03} and also found 
a groove-like structure, however, the poor statistics is not sufficient to constrain the parameters of such a resonance.
Therefore, three possibilities arise: 
(i) the data reduction procedure might have some problems in Ref.~\cite{bib:gom01}. They had to reconstructed the excitation 
functions (above 2.1 MeV) with some technical treatment since their high-energy protons escaped from two thin Si detectors;
(ii) the peak observed in Ref.~\cite{bib:gom01} may be due to the inelastic-scattering contribution~\cite{bib:bar12}, or the carbon 
(from (CH$_{2}$)$_{n}$ target itself) induced background which was not measured and subtracted accordingly;
(iii) the 1$^-$ assignment for the 6.15-MeV state was wrong in Ref.~\cite{bib:gom01}. If their data were correct, it shows that 
the 6.15-MeV state most probably has a 3$^-$ or 2$^-$ assignment, while the 6.30-MeV state is the key 1$^-$ state~\cite{bib:he10}. 
In this case, the calculated reaction rates for $^{14}$O($\alpha$, $p$)$^{17}$F are quite different from the previous 
estimations~\cite{bib:he10}. In order to clarify the above discrepancies, we have performed a new $^{17}$F($p$,$p$)$^{17}$F elastic 
scattering measurement with a thick-target method~\cite{bib:art90,bib:gal91,bib:kub01}, which proved to be a 
successful technique in our previous studies~\cite{bib:ter03,bib:ter07,bib:hjj07,bib:yam09,bib:he09}.

The experiment was performed using the CNS radioactive ion beam separator (CRIB)~\cite{bib:yan05,bib:kub02}, installed by the Center 
for Nuclear Study (CNS), University of Tokyo, in the RIKEN Accelerator Research Facility. A primary beam of $^{16}$O$^{6+}$ was 
accelerated up to 6.6 MeV/nucleon by an AVF cyclotron ($K$=70) with an average intensity of 560 enA. The primary beam bombarded a 
liquid-nitrogen-cooled D$_{2}$ gas target (at 90 K)~\cite{bib:yam08} where the radioactive ion (RI) beam of $^{17}$F was produced via 
the $^{16}$O($d$,$n$)$^{17}$F reaction in inverse kinematics. The D$_{2}$ gas in 120 Torr pressure was confined in a 80-mm long cell 
with two 2.5 ${\mu}$m thick Havar foils. The $^{17}$F beam was separated by the CRIB separator and identified by using a in-flight 
method. The $^{17}$F beam, with a mean energy of 61.9$\pm$0.5 MeV and an average intensity of 2.5${\times}$10$^{5}$ pps, was 
bombarded a thick H$_{2}$ gas target in an F3 chamber. The beams were stopped completely in this target.

\begin{figure}
\begin{center}
\includegraphics[scale=0.35]{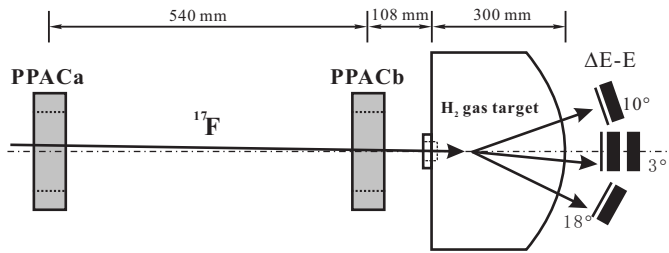}
\end{center}
\caption{\label{prcfig:fig1} Schematic diagram of the experimental setup at F3 chamber, similar to that used in Ref.~\cite{bib:korea}.}
%\label{fig1}
\end{figure}

\begin{figure}
\includegraphics[scale=0.5]{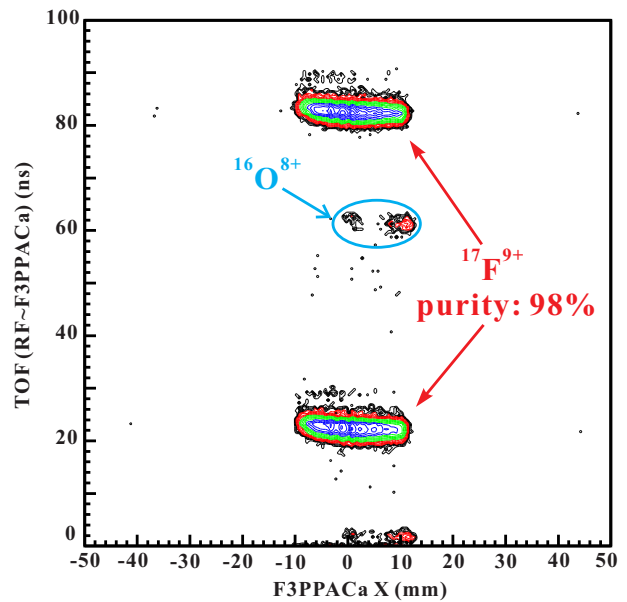}
\caption{\label{prcfig:fig2} Identification plot for the beam particles before target.}
\end{figure}

The experimental setup at F3 chamber is shown in Fig.~\ref{prcfig:fig1}, which is quite similar to that used in Ref.~\cite{bib:korea}.
The beam purity was about 98\% after passing through a Wien-filter, which is capable of purifying the secondary beam efficiently.
Two PPACS (Parallel Plate Avalanche Counters)~\cite{bib:kum01} were used for measuring time and two-dimensional position information 
of the beam particles. The beam profile on the secondary target was monitored by the PPACs during the data acquisition. The beam 
particles were identified in an event-by-event mode by using the abscissa of PPACa, and the TOF between PPACa and the RF signal 
provided by the cyclotron. Fig.~\ref{prcfig:fig2} shows the particle identification just before the secondary target.
The H$_{2}$ gas target in a 600 Torr pressure was housed in a 300-mm-radius semi-cylindrical shape chamber sealed with a 
2.5-$\mu$m-thick Havar foil as a beam entrance window and with a 25-$\mu$m-thick aluminized Mylar foil as an exit window. Comparing 
to the solid polyethylene (CH$_{2}$)$_{n}$ target, the gas target is free from intrinsic background contributions.

The recoiled light particles were measured by using three sets of ${\Delta}$E-E Si telescope at averaged angles of 
$\theta$$_{lab}$$\approx$3$^\circ$, 10$^\circ$ and 18$^\circ$, respectively. In the {\it c.m.} frame of elastic scattering, the 
relevant averaged scattering angles are determined to be $\theta_{c.m.}$$\approx$155$^\circ$$\pm$18$^\circ$, 
138$^\circ$$\pm$20$^\circ$ and 120$^\circ$$\pm$22$^\circ$, respectively. At $\theta$$_{lab}$$\approx$3$^\circ$, the telescope was 
consisted of a 65-$\mu$m-thick W1-type double-sided-strip 
(16$\times$16 strips) silicon detector and two 1500-$\mu$m-thick MSX25-type pad detectors. The last pad detector was acted as a veto 
to eliminate the high energy proton background. The configuration of other two telescopes is similar to that at 
$\theta$$_{lab}$$\approx$3$^\circ$, just no third veto detectors. The position sensitive $\Delta E$ detectors measured the energy, 
position and timing signals of the particles, and the pad $E$ detectors measured their residual energies. The recoiled particles were 
clearly identified by using a ${\Delta}$E-E method. The energy calibration for the silicon detectors was performed by using a 
standard triple ${\alpha}$ source, and the secondary proton beams at several energy points. In addition, an Ar gas at 120 Torr filled 
in the gas target chamber, was used in a separate run for evaluating the background contribution.

For inverse kinematics, the center-of-mass ({\it c.m.}) energy $E_{c.m.}$ of the $^{17}$F+$p$ system is related to the energy $E_p$ 
of the recoil protons detected at a laboratory angle $\theta_{lab}$ by~\cite{bib:he07}
\begin{eqnarray}
E_{c.m.}=\frac{A_b + A_t}{4 A_b \mathrm{cos}^2 \theta_{lab}} E_p
\label{eq:one},
\end{eqnarray}
where $A_b$ and $A_t$ are the mass numbers of the beam and target nuclei. This equation is valid only for an elastic scattering case. 
Practically, $E_{p}$ was converted to $E_{c.m.}$ by assuming the elastic scattering kinetics and considering the energy loss of 
particles in the target. 
The laboratory differential cross sections ($d\sigma$/$d\Omega$) for $^{17}$F+$p$ elastic scattering at energy $E_p$ and angle 
$\theta_{lab}$ are deduced from the proton spectrum by the following equation~\cite{bib:kub01,bib:he07}
\begin{eqnarray}
\frac{d\sigma}{d\Omega_{lab}}(E_p,\theta_{lab})=\frac{N}{I_0 N_s \Delta\Omega_{lab}}
\label{eq:two},
\end{eqnarray}
Where $N$ is the number of detected protons, {\em i.e.}, at energy interval of $E_p \rightarrow E_p+ \Delta E$ and scattering angle 
of $\theta_{lab}$, which are measured by a Si telescope covering a solid angle $\Delta\Omega_{lab}$. $I_0$ is the total number of 
$^{17}$F beam particles bombarded the $H_2$ target, and it is considered to be constant in the whole energy region. $N_s$ is
the number of H atoms per unit area per energy bin in the target (${dx/dE}$)~\cite{bib:zie85}. The transformation of the laboratory 
differential cross sections to the {\it c.m.} frame is given by~\cite{bib:he07}
\begin{eqnarray}
\frac{d\sigma}{d\Omega_{c.m.}}(E_{c.m.},\theta_{c.m.})=\frac{1}{4 \mathrm{cos} \theta_{lab}} \frac{d\sigma}{d\Omega_{lab}}(E_{p},\theta_{lab})
\label{eq:three}.
\end{eqnarray}
Unlike thin solid target, the gas target has a length of 300 mm, and hence one has to take care of the accuracy of reaction-position 
determination. The uncertainty of the reaction position can cause the error of the solid angle ${\Delta\Omega_{lab}}$ ($\sim$3\%), 
which is the main error in cross-section data other than the statistical one ($\sim$1\%).

\begin{figure}
\includegraphics[scale=0.45]{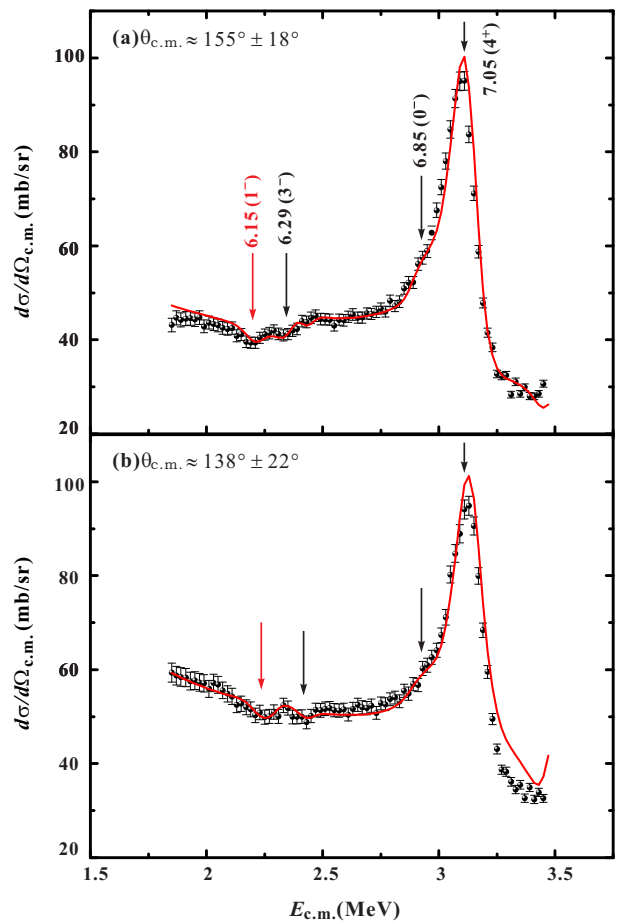}
\caption{\label{prcfig:fig3} The {\it c.m.} differential cross-sections for elastically scattered protons produced by bombarding a thick
H$_{2}$ gas target with a 61.9 MeV $^{17}$F beam at scattering angles of (a) $\theta$$_{c.m.}$$\approx$155$^\circ$$\pm$18$^\circ$
and (b) $\theta$$_{c.m.}$$\approx$138$^\circ$$\pm$22$^\circ$. The curve lines represent the $R$-matrix fits in which a high-lying
resonance ($E_x$=7.35 MeV, $J^\pi$=2$^+$) was involved.}
\end{figure}

The reconstructed excitation functions of $^{17}$F+$p$ at two scattering angles are shown in Fig.~\ref{prcfig:fig3}. The background 
was subtracted with the data taken from the Ar gas run.
Several resonance structures were clearly observed. In order to determine the resonant parameters of observed resonances, the 
multichannel $R$-matrix calculations~\cite{bib:lan58,bib:des03,bib:bru02} (see examples~\cite{bib:he10,bib:mur09}) that include the 
energies, widths, spins, angular momenta, and interference sign for each candidate resonance have been performed in the present work. 
A channel radius of $R$=1.25$\times$(1+17$^{\frac{1}{3}}$) fm appropriate for the 
$^{17}$F+$p$ system~\cite{bib:wie87,bib:hah96,bib:har02} has been utilized in the calculation. The ground-state spin-parity 
configurations of $^{17}$F and proton are $5/2^{+}$ and $1/2^{+}$, respectively. Thus, there are two channel spins in the elastic 
channel, {\it i.e.}, $s$=2 or 3.

Four resonances, {\it i.e.}, at $E_{x}$=6.15, 6.30, 6.85, and 7.05 MeV, have been analyzed and the most probable fitting curves are 
shown in Fig.~\ref{prcfig:fig3}. It shows that the fitting results at different scattering angles are well consistent with each other.
According to our $R$-matrix analysis, a groove-like structure around $E_{c.m.}$=2.21 MeV corresponding to the 6.15 MeV state in 
$^{18}$Ne can be fitted as a 1$^-$ ($s$=2, $\ell$=1), $\Gamma_p$=50 keV very well. This resonance shape is totally different from 
the bump-like shape observed in Ref.~\cite{bib:gom01}. Thus, we now think that 1$^-$ assignment to the key 6.15 MeV state is verified 
on a much firmer experimental ground, and it is definitely the key resonance in calculating the $^{14}$O($\alpha$,$p$)$^{17}$F reaction 
rate below 1.0 GK.
%%%%%%%%%%%%%%
An obvious structure at $E_{x}$=6.29 MeV was observed in the excitation function, and its shape is reproduced very well with 
Hahn {\it et al.}~\cite{bib:hah96} resonant parameters, {\it i.e.}, $E_{c.m.}$=2.37~MeV, $J^\pi$=3$^-$, and $\Gamma_p$=20 keV. 
In Ref.~\cite{bib:gom01}, this state was not clearly observed and hence not involved in their $R$-matrix fit.
%%%%%%%%%%%%%%
Moreover, a shoulder-like structure around $E_{c.m.}$=2.93 MeV was clearly observed as shown in Fig.~\ref{prcfig:fig3}. This should 
be a new state at $E_{x}$=6.85 MeV. According to the $R$-matrix analysis, it is probably a $J^\pi$=0$^-$ state 
($\Gamma_p$$\sim$50 keV). This new state is most likely the analog state of $^{18}$O at $E_x$=6.880 MeV. The energy shift of 
30 keV is consistent with the small energy shift for the negative-parity states in this excitation energy 
region~\cite{bib:for00}.
%%%%%%%%%%%%%%
The well-known state~\cite{bib:har02} at $E_{x}$=7.05 MeV (4$^+$, $\Gamma_p$$\sim$95 keV) was also observed at $E_{c.m.}$=3.13 MeV 
in the excitation function. However, the doublet structure around $E_{x}$=7.05 and 7.12 MeV suggested in 
Refs.~\cite{bib:hah96,bib:he11} could not be resolved within the present energy resolution ($\sim$80 keV).

In a summary, four resonances in $^{18}$Ne, {\it i.e.}, at $E_{x}$=6.15, 6.30, 6.85, and 7.05 MeV, have been observed clearly and 
analyzed in the present work. Especially, we made a firm $J^\pi$=1$^-$ assignment for the key state at 6.15 MeV of great astrophysical 
interest. The confirmative spin-parity 1$^{-}$ assignment for this state clarifies the discrepancies arising in the literature. 
In addition, a new state observed at $E_{x}$=6.85 MeV was tentatively assigned as $J^\pi$=0$^{-}$, which could be the mirror state of 
6.880 MeV, 0$^{-}$ in $^{18}$O. The data analysis of the two-proton emission from the resonances and the measured ($p$,$\alpha$) 
cross section data are still under progress, and the results will be published elsewhere~\cite{bib:hu13}.

We would like to thank the RIKEN and CNS staff for their friendly operation of the AVF cyclotron. This work is financially supported 
by the National Natural Science Foundation of China (Nos. 11135005, 11021504), the Major State Basic Research Development Program of 
China (2013CB834406), as well as supported by the JSPS KAKENHI (No. 21340053).


\begin{thebibliography}{99}
\bibitem{bib:cha92}
A.E. Champage and M. Wiescher, Annu. Rev. Nucl. Part. Sci. \textbf{42}, 39 (1992).
\bibitem{bib:wie99}
M. Wiescher {\it et al.}, J. Phys. G \textbf{25}, R133 (1999).
\bibitem{bib:sch01}
H. Schatz {\it et al.}, Phys. Rev. Lett. \textbf{86}, 3471 (2001).
\bibitem{bib:wan12}
M. Wang {\it et al.}, Chin. Phys. C \textbf{36}(12), 1603 (2012).
\bibitem{bib:hah96}
K.I. Hahn {\it et al.}, Phys. Rev. C \textbf{54}, 1999 (1996).
\bibitem{bib:gom01}
J. G\'omez del Campo {\it et al.}, Phys. Rev. Lett. \textbf{86}, 43 (2001).
\bibitem{bib:bla03}
J.C. Blackmon {\it et al.}, Nucl. Phys. A \textbf{718}, 127(c) (2003).
\bibitem{bib:cern}
J.J. He {\it et al.}, Phys. Rev. C \textbf{80}, 042801(R) (2009).
\bibitem{bib:bar10}
D.W. Bardayan {\it et al.}, Phys. Rev. C \textbf{81}, 065802 (2012).

\bibitem{bib:not04}
M. Notani {\it et al.}, Nucl. Phys. A \textbf{746}, 113(c) (2004).
\bibitem{bib:kub06}
S. Kubono {\it et al.}, Eur. Phys. J. A \textbf{27}, 327 (2006).
\bibitem{bib:bla01}
J.C. Blackmon {\it et al.}, Nucl. Phys. A \textbf{688}, 142(c) (2001).
\bibitem{bib:har99}
B. Harss {\it et al.}, Phys. Rev. Lett. \textbf{82}, 3964 (1999).
\bibitem{bib:har02}
B. Harss {\it et al.}, Phys. Rev. C \textbf{65}, 035803 (2002).

\bibitem{bib:wie87}
M. Wiescher {\it et al.}, Astrophys. J. \textbf{316}, 162 (1987).
\bibitem{bib:he10}
J.J. He {\it et al.}, arXive:1001.2053v1.
\bibitem{bib:he11}
J.J. He {\it et al.}, Eur. Phys. J. A \textbf{47}, 67 (2011).
\bibitem{bib:bar12}
D.W. Bardayan {\it et al.}, Phys. Rev. C \textbf{85}, 065805 (2012).

\bibitem{bib:art90}
K.P. Artemov {\it et al.}, Sov. J. Nucl. Phys. \textbf{52}, 408 (1990).
\bibitem{bib:gal91}
W. Galster {\it et al.}, Phys. Rev. C \textbf{44}, 2776 (1991).
\bibitem{bib:kub01}
S. Kubono, Nucl. Phys. \textbf{A693}, 221 (2001).

\bibitem{bib:ter03}
T. Teranishi {\it et al.}, Phys. Lett. B \textbf{556}, 27 (2003).
\bibitem{bib:ter07}
T. Teranishi {\it et al.}, Phys. Lett. B \textbf{650}, 129 (2007).
\bibitem{bib:hjj07}
J.J. He {\it et al.}, Phys. Rev. C \textbf{76}, 055802 (2007).
\bibitem{bib:yam09}
H. Yamaguchi {\it et al.}, Phys. Lett. B \textbf{672}, 230 (2009).
\bibitem{bib:he09}
J.J. He {\it et al.}, Phys. Rev. C \textbf{80}, 015801 (2009).

\bibitem{bib:yan05}
Y. Yanagisawa {\it et al.}, Nucl. Instrum. Meth. A \textbf{539}, 74 (2005).
\bibitem{bib:kub02}
S. Kubono {\it et al.}, Eur. Phys. J. A \textbf{13}, 217 (2002).
\bibitem{bib:yam08}
H. Yamaguchi {\it et al.}, Nucl. Instr. Meth. A \textbf{589}, 150 (2008).
\bibitem{bib:korea}
H.S. Jung {\it et al.}, Phys. Rev. C \textbf{85}, 045802 (2012).
\bibitem{bib:kum01}
H. Kumagai {\it et al.}, Nucl. Instr. Meth. A \textbf{470}, 562 (2001).
\bibitem{bib:kub01}
S. Kubono, Nucl. Phys. A \textbf{693}, 221 (2001).

\bibitem{bib:zie85}
J.F. Ziegler {\it et al.}, {\it The Stopping and Range of Ions in Solids} (Pergamon Press, New York, 1985).
\bibitem{bib:lan58}
A.M. Lane and R.G. Thomas, Rev. Mod. Phys. \textbf{30}, 257 (1958).
\bibitem{bib:des03}
P. Descouvemont, {\it Theoretical Models for Nuclear Astrophysics} (Nova Science Publishers Inc., New York, 2003).
\bibitem{bib:bru02}
C.R. Brune, Phys. Rev. C \textbf{66}, 044611 (2002).
\bibitem{bib:mur09}
A.St.J. Murphy {\it et al.}, Phys. Rev. C \textbf{79}, 058801 (2009).
\bibitem{bib:hu13}
J. Hu {\it et al.}, under preparation.
\bibitem{bib:for00}
H.T. Fortune and R. Sherr, Phys. Rev. Lett. \textbf{84}, 1635 (2000).
%\bibitem{bib:for03}
%H.T. Fortune and R. Sherr, Phys. Rev. C \textbf{68}, 034307 (2003).

\end{thebibliography}
\end{document}